\documentclass[amsmath,amssymb,prl,twocolumn]{revtex4}
\usepackage{graphicx}
\usepackage{dcolumn}
\usepackage{bm}
\begin{document}

\title{Conduction properties of extended defect states in Dirac materials}

\author{Francesco Romeo}
\affiliation{Dipartimento di Fisica "E.R. Caianiello", Universit\`a di Salerno, I-84084 Fisciano (Sa), Italy}

\date{\today}
\begin{abstract}
We demonstrate the existence of localized states in close vicinity of a linear defect in graphene. These states have insulating or conducting character. Insulating states form a flat band, while conducting states present a slowdown of the group velocity which is not originated by many-body interactions and it is controlled by the interface properties. For appropriate boundary conditions, the conducting states exhibit momentum-valley locking and protection from backscattering effects. These findings provide a contribution to the recent discussion on the origin of correlated phases in graphene.
\end{abstract}


\maketitle
In his seminal work \cite{feynman82} R. Feynman proposed the striking idea that the dynamics of a given quantum system can be simulated by the evolution of a second one. This amazing statement, which introduces the concept of analog model, is actually in logical continuity with the simulation of mechanical systems by using the early analog computers \cite{fpu55}. The modern development of these concepts is leading to manmade quantum simulators and quantum computers \cite{quantumcomp96,dwavepc16}.\\
Recently, nature-given non-relativistic quantum systems which are able to emulate ultra-relativistic behaviour have been discovered. These physical systems are the so-called Dirac materials of which graphene is the prototype \cite{dassarma2011}. Interestingly, studies in photonic crystals and optical metamaterials are nowadays inspired by the graphene structure \cite{polini13}.\\
Graphene is a two-dimensional honeycomb lattice of carbon atoms. The hexagonal Brillouin zone has six corners (K/K' points) in the vicinity of which the low-energy region of the band's structure is described by the Dirac equation. In recent past, graphene has attracted considerable attention for its properties, such as for instance Klein tunneling \cite{kleintun}, Zitterbewegung effect \cite{rusin08}, antilocalization \cite{tikhonenko09}, anomalous quantum Hall effect \cite{ostrovsky08} and Veselago focusing effect \cite{cheianov07}, which in many cases have been also experimentally confirmed.\\
Interest in graphene has recently found renewed impetus motivated by the discovery of superconducting and Mott phases in twisted bilayer and trilayer graphene \cite{cao18a,cao18b,chen19}. These phases have been observed when charge carriers partly fill a nearly flat band which appears at specific twist angles (magic angles) \cite{magicangles19}. The existence of a flat band suggests that renormalization of particles velocity plays a relevant role although it is unclear whether particles velocity slowdown is the cause or effect of the emergence of correlated phases in graphene.\\
\begin{figure}[!t]
\includegraphics[clip,scale=0.5]{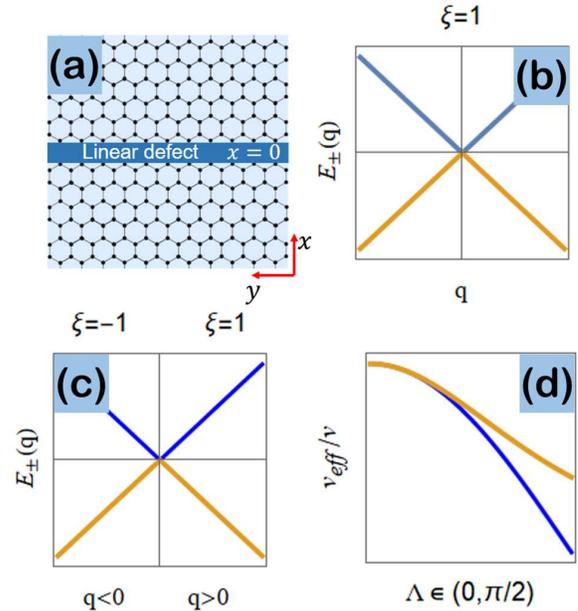}{\centering}\\
\caption{(color online). (a) Linear defect with translational invariance along the y-direction. (b) Linear dispersion relation of conducting states with interface potential $V(x,y)=\hbar v \mathbb{I}\Lambda \delta(x)$. (c) Linear dispersion relation of conducting states with valley-momentum locking obtained for $V(x,y)=\hbar v \sigma_{y}\Lambda \delta(x)$. (d) Renormalized propagation velocity $v_{eff}$ as a function of the interface parameter $\Lambda$. Upper (lower) curve is obtained for $v_{eff}=v/\cosh(\Lambda)$ ($v_{eff}=v |\cos(\Lambda)|$), with $v$ the bulk propagation velocity.}
\label{fig:fig1}
\end{figure}
On the other hand, the existence of localized states in close vicinity of graphene grain boundaries has been experimentally demonstrated \cite{gbstates16}; the conductance of these states is reduced compared to bulk and this suggests that a slowdown of the group velocity of localized modes also occurs in a graphene-based system with a simpler geometry with respect to the bilayer graphene.\\
Motivated by these observations, in this Letter we theoretically demonstrate the existence of localized states in close vicinity of a linear defect of a graphene monolayer, which is here considered as a paradigmatic realization of a Dirac material. These states have insulating or conducting character and present a slowdown of the group velocity which is controlled by the interface properties. The velocity renormalization described in this work is not originated, as usual \cite{giamarchibook}, by many-body interactions, while it is a single particle effect which can eventually drive the system towards a correlated phase. For appropriate boundary conditions, the conducting states exhibit protection from backscattering effects due to momentum-valley locking which prevents changing the momentum orientation without simultaneously change the valley quantum number.\\
To follow our program, we present a minimal model of a graphene monolayer which includes a defects line with translational invariance along the $y$-direction (Fig. \ref{fig:fig1} (a)). The two-valley Dirac Hamiltonian of the mentioned problem takes the following form \cite{note}:
\begin{equation}
\label{eq:hamiltonian}
\mathcal{H}=\sum_{\xi= \pm 1} K_{\xi}\otimes H_{\xi}+U(x,y),
\end{equation}
where the $2 \times 2$ matrix $K_{\xi}=(\mathbb{I}+\xi \sigma_z)/2$ defines the valley subspace, while $H_{\xi}=v(\xi \sigma_x p_x+\sigma_y p_y)$ represents the Hamiltonian describing spinless particles with assigned valley quantum number $\xi=\pm 1$ and propagation velocity $v$. The spinorial structure of the theory is not related with the particles spin (which is here neglected) but is reminiscent of the two-atoms structure of the graphene unit cell. The effect of a linear defect is modeled by including the single-particle potential $U(x,y)=\sum_{\xi} K_{\xi} \otimes \mathcal{B}_{\xi} \delta(x)$, with $\mathcal{B}_{\xi}=\mathcal{B}^{\dag}_{\xi}$. Under the long-wavelength limit of the particles dynamics which is the underlying assumption when a continuous treatment is adopted, the Dirac delta potential profile is justified when particles wavelength is much greater than the defect extension along the $x$-direction. We also assume that intervalley scattering events, if present, are infrequent and thus a diagonal potential with respect to the valley quantum number can be safely considered. Once the relevant time-reversal operator $\mathcal{T}=(\sigma_{x} \otimes \mathbb{I})\mathcal{K}$ is introduced as done in \cite{trgraphene}, the requirement of preserving the time reversal symmetry (i.e. $[H,\mathcal{T}]=0$) implies $\mathcal{B}^{\ast}_{\xi}=\mathcal{B}_{-\xi}$. Under the above assumptions, the description of particles belonging to different valleys can be treated separately and thus, without loss of generality, we set $\xi=+1$ and look for a localized eigenstate of the Hamiltonian $H_{+}+\mathcal{B}_{+} \delta(x)$, the specific structure of $\mathcal{B}_{+}$ being left unspecified for the moment. By definition, a localized state cannot carry charge current along the $x$-direction and this constrains the spinorial components of such a state. The above requirement does not exclude that the localized state might behave like a current-carrying state along the $y$-direction.\\
We first demonstrate the existence of insulating states completely delocalized along the defect line for which the expectation values of the current density operator along the $x$- and $y$-direction, namely $J_{x/y}=v \sigma_{x/y}$, are both vanishing quantities. Translational invariance requires that an eigenstate of the problem takes the form $\Psi(x,y)=(\Phi_{A}(x),\Phi_{B}(x))^{t}\exp(i q y)$, while the additional conditions $\Psi^{\dag}(x,y)\sigma_{x/y}\Psi(x,y)=0$ imply the following relations among the spinorial components:
\begin{eqnarray}
\label{eq:spininscond}
&&\Phi^{\ast}_{A}(x)\Phi_{B}(x)+\Phi^{\ast}_{B}(x)\Phi_{A}(x)=0\nonumber\\
&&\Phi^{\ast}_{B}(x)\Phi_{A}(x)-\Phi^{\ast}_{A}(x)\Phi_{B}(x)=0.
\end{eqnarray}
A general solution of Equations (\ref{eq:spininscond}) is obtained by requiring that only one component of the spinor is different from zero. Direct solution of Dirac equation for $x\neq 0$ shows that the only way to get non-trivial solution is assuming a dispersion relation $E_{q}=0$, meaning that if such a state exists it defines a flat band at the Dirac point. Fixing for deﬁniteness $q>0$, $\Psi(x,y)=\exp(i q y)\exp(-q |x|)[\Phi_{+}\theta(x)+\Phi_{-}\theta(-x)]$ represents a translational-invariant solution written in terms of the Heaviside step function $\theta(x)$. The spinors defining the wavefunction take the form $\Phi_{+}=(0,\varphi_{+})^{t}$ and $\Phi_{-}=(\varphi_{-},0)^{t}$ with the constants $\varphi_{\pm}$ being fixed by the boundary conditions at $x=0$ (imposed by the interface potential) and by the normalization. In general, it is easy to show that $\Phi^{\dag}_{+}\Phi_{-}=0$ is a necessary requirement to get translational invariant solutions. Boundary conditions at $x=0$ are implemented by using the matching matrix method in the form $\Psi(0^{+},y)=\mathcal{M}\Psi(0^{-},y)$, with the matching matrix $\mathcal{M}$ directly derived from the interface potential according to the relation $\mathcal{M}=\exp(-i \sigma_{x}\mathcal{B}_{+}/(\hbar v))$ \cite{matchingm}. Inspection of the boundary conditions shows that the quantum state we are talking about nucleates at the interface provided that $\mathcal{M}_{11}=0$, i.e. for appropriate interface potentials. As an instructive example, it is easily proven that the interface potential $V(x,y)=\hbar v \mathbb{I}\Lambda \delta(x)$ admits a matching matrix $\mathcal{M}=\cos(\Lambda)\mathbb{I}-i \sigma_{x}\sin(\Lambda)$ which supports the mentioned states when $\cos(\Lambda)=0$, with $\Lambda$ a dimensionless parameter. On the other hand, $V(x,y)=\hbar v \sigma_{z}\Lambda \delta(x)$, with $\mathcal{M}=\cosh(\Lambda)\mathbb{I}-\sigma_{y}\sinh(\Lambda)$, does not support insulating states since $\mathcal{M}_{11}=\cosh(\Lambda)>0$.\\
Conductive states can be obtained requiring the confinement condition along the $x$-direction without additional constraints. Direct computation shows that the quantum state $\Psi(x,y)=(i\phi_{A}(x),\phi_{B}(x))^{t}\exp(i q y)$, with $\phi_{A/B}(x)$ real valued auxiliary functions, fulfills the confinement condition. Substituting the trial wavefunction $\Psi(x,y)$ into the Dirac equation and after straightforward algebra one easily get decoupled differential equations $\partial^{2}_{x}\phi_{\alpha}(x)+(\epsilon^2-q^2)\phi_{\alpha}(x)=0$ ($\alpha \in\{A,B\}$) for the auxiliary functions. Equations depend on the linear momentum $q$ along the $y$-direction and on $\epsilon=E/(\hbar v)$, which is a quantity related to the energy eigenvalue $E$. The solutions of the differential equations are linear combination of exponentials $\exp(\pm \sqrt{q^2-\epsilon^2})$ for which the confinement condition requires $q^2-\epsilon^2>0$, while the arbitrary coefficients of the linear combination are constrained by the asymptotic behavior of the wavefunction, i.e. $lim_{x\rightarrow \pm \infty}\Psi(x,y)=0$. Considering all constraints, we obtain the following expression for the spinorial wavefunction:
\begin{eqnarray}
\label{eq:spinorcondst}
&&\Psi(x,y)=\\
&&\sum_{s=\pm}\theta(s x)\mathcal{A}_{s}\exp(i q y)\exp(-\lambda |x|)\left[
                                                                                       \begin{array}{c}
                                                                                         -i(\frac{q-s \lambda}{\epsilon}) \\
                                                                                         1 \\
                                                                                       \end{array}
                                                                                     \right]\nonumber,
\end{eqnarray}
where we have introduced the decay length $\lambda=\sqrt{q^2-\epsilon^2}$. Equation (\ref{eq:spinorcondst}) represents a well-defined spinor provided that the conditions $\lambda>0$ and $\epsilon \neq 0$ are respected. The above requirements can be satisfied or not depending on the boundary conditions which, together with the normalization condition, allows us to find the unknown coefficients $\mathcal{A}_{\pm}$ and $\epsilon$, the latter being related to the energy eigenvalue. When the boundary conditions in $x=0$ are taken into account within the matching matrix method discussed before, the following spectrum equation is derived:
\begin{eqnarray}
\label{eq:spectrum}
q(\mathcal{M}_{11}-\mathcal{M}_{22})&+&\sqrt{q^2-\epsilon^2}(\mathcal{M}_{11}+\mathcal{M}_{22})+\nonumber\\
&+&i(\mathcal{M}_{12}+\mathcal{M}_{21})\epsilon=0,
\end{eqnarray}
to be solved with respect to $\epsilon$ under the constraints $\epsilon \neq 0$ and $q^2-\epsilon^2\neq 0$. Equation (\ref{eq:spectrum}) immediately implies that the interface potential $V(x,y)=\hbar v \sigma_{z}\Lambda \delta(x)$, with $\mathcal{M}_{11}=\mathcal{M}_{22}=\cosh(\Lambda)$ and $\mathcal{M}_{12}=-\mathcal{M}_{21}=i\sinh(\Lambda)$, does not support conducting states since no acceptable solution of the spectrum equation exists. We reach the same conclusion when considering the potential $V(x,y)=\hbar v \sigma_{x}\Lambda \delta(x)$, with $\mathcal{M}=\exp(-i\Lambda)\mathbb{I}$.\\
On the other hand, the interface potential $V(x,y)=\hbar v \mathbb{I}\Lambda \delta(x)$ generates matching conditions ($\mathcal{M}_{11}=\mathcal{M}_{22}=\cos (\Lambda)$ and $\mathcal{M}_{12}=\mathcal{M}_{21}=-i\sin(\Lambda)$) that allow to get solutions of the spectrum equation. In particular, we find that the dispersion relation $E_{\nu}(q)=\nu \hbar v_{eff}|q|$ (Fig. \ref{fig:fig1} (b)), with particle-hole index $\nu=\pm 1$ and effective velocity
\begin{equation}
\label{eq:renvel1}
v_{eff}=v |\cos(\Lambda)|,
\end{equation}
is solution of the spectrum equation provided that $sign(\cos(\Lambda))+\nu sign(\sin(\Lambda))=0$, $\cos(\Lambda)\neq 0$ and $\sin(\Lambda)\neq 0$. Once the interface parameter $\Lambda$ has been assigned, states with positive (electron-like) or negative (hole-like) energy are stabilized depending on the sign of $\nu$ which is determined by
\begin{equation}
\nu=-\frac{sign(\cos(\Lambda))}{sign(\sin(\Lambda))}.
\end{equation}
The physical meaning of the above relation is that the considered potential cannot simultaneously have confining properties for particles belonging to the conduction ($\nu=+1$) and valence ($\nu=-1$) band; thus only one species at the time is confined by the potential once the interface parameter $\Lambda$ has been specified.
The conducting states have one dimensional character and present a group velocity $v_g=\hbar^{-1}\partial_{q}E_{\nu}(q)= \nu \ v_{eff} sign(q)$, being the sign of the linear momentum $q$ either positive or negative. Interestingly, conductive states present a renormalization of the propagation velocity (i.e., $v_{eff}<v$) completely determined by the interface properties.\\
So far we have found conducting states with an interface potential which is known to preserve the Klein tunneling. However, graphene grain boundaries are a relevant example of linear defect where Klein tunneling is suppressed with a concomitant conductance reduction of the graphene/grain-boundary/graphene junction. A minimal model of the situation mentioned above can be obtained by considering the interface potential $V(x,y)=\hbar v \sigma_{y}\Lambda \delta(x)$ which leads to a matching matrix belonging to the linear special group SL(2,$\mathbb{C}$) with elements $\mathcal{M}_{11}=\exp(\Lambda)$, $\mathcal{M}_{22}=\exp(-\Lambda)$ and  $\mathcal{M}_{12}=\mathcal{M}_{21}=0$. Specializing Equation (\ref{eq:spectrum}) to the present case and after straightforward algebra one get:
\begin{equation}
\label{eq:locking}
\sqrt{q^2-\epsilon^2}=q \Big(\frac{1-\alpha^2}{1+\alpha^2}\Big),
\end{equation}
with $\alpha=\exp(\Lambda)$. Equation (\ref{eq:locking}) is an equivalent form of the spectrum equation explicitly showing that $sign(q)$ must be equal to $sign(1-\alpha^2)$ in order to ensure that a solution of the spectrum equation exists. Thus, fixing the valley quantum number $\xi=+1$, the orientation of the current flow is determined by the interface properties and a bidirectional conducting channel along the interface cannot be obtained with quantum states belonging to the same valley. The latter statement immediately follows observing that if a current-carrying state with momentum $q$ and $\xi=+1$ exists, it is not accompanied by a state with quantum numbers $-q$ and $\xi=+1$. Solving Equation (\ref{eq:locking}) with respect to $\epsilon$ leads to the dispersion relation $E_{\nu}(q)=\nu \hbar v_{eff}|q|$ (Fig. \ref{fig:fig1} (c)) with effective velocity given by
\begin{equation}
v_{eff}=\frac{v}{\cosh(\Lambda)},
\end{equation}
with $v_{eff}<v$. The group velocity associated with the conducting states is given by $v_g=\hbar^{-1}\partial_{q}E_{\nu}(q)= \nu \ v_{eff} sign(1-\alpha^2)$, while the spinorial wavefunction takes the form:
\begin{eqnarray}
\label{eq:spinorbis}
&&\Psi^{(\xi=+1)}_{\nu}(x,y)=\\
&&\left\{
              \begin{array}{ll}
                \mathcal{A}\exp(i q y)\exp(-\lambda |x|)\left[
                                                          \begin{array}{c}
                                                            -i \nu sign(q) \\
                                                            \alpha \\
                                                          \end{array}
                                                        \right] & \hbox{$x<0$} \\
\\
               \mathcal{A}\exp(i q y)\exp(-\lambda |x|)\left[
                                                         \begin{array}{c}
                                                          -i \nu sign(q) \alpha \\
                                                           1 \\
                                                         \end{array}
                                                       \right] & \hbox{$x>0,$}\nonumber
              \end{array}
            \right.
\end{eqnarray}
with $\mathcal{A}$ a normalization constant, $q=|q|sign(1-\alpha^2)$, $\lambda=|q| |1-\alpha^2|/(1+\alpha^2)$ and $\alpha \neq 1$. For opaque interfaces, which are described by $\alpha\gg 1$, the approximate relation $\lambda \approx |q|$ holds. The analysis of the same problem for particles with valley quantum number $\xi=-1$ shows that $\Psi^{(\xi=-1)}_{\nu}(x,y)$ takes the same structure of $\Psi^{(\xi=1)}_{\nu}(x,y)$ even though in that case $q=-|q|sign(1-\alpha^2)$. The above observation implies that these quantum states define a bidirectional ballistic conducting channel protected from backscattering effects. This protection is originated by the momentum-valley locking mechanism which prevents the sign reversal of linear moment $q$ without the simultaneous change of the valley quantum number $\xi$. Moreover, following \cite{matchingm}, it can be shown that the interface potential of a grain boundary junction with rotated crystallographic axes retains a dependence on the misorientation angle. According to this observation, we expect that in graphene grain boundaries the renormalized modes velocity $v_{eff}$ can be controlled by the misorientation angle between the crystallographic axes of the two sides of the junction.\\
Accordingly, it is argued that conductive states in close vicinity of linear defect in graphene (such as a grain boundary) can act as nucleation center of emergent correlated phases. Indeed, conductive states pinned at the defect have a reduced kinetic energy compared to the bulk states and thus they are exposed to the effects of particle-particle interaction. In principle, under appropriate conditions, hybridization between bulk and defect states can facilitate the diffusion of a correlated phase in the vicinity of the defect. This mechanism can be tested and exploited by creating engineered defects also using the proximity effect induced by superconducting materials. Graphene-superconductor proximity coupling \cite{beenakker06} represents a natural way to get communication between particles belonging to different valleys, the latter being an interesting resource to explore exotic states of matter. Going beyond condensed matter systems, selected aspects of the mentioned phenomenology can be studied in photonic metamaterials where the analog of graphene can be obtained \cite{photonicgraphene}.

In conclusion, we have demonstrated that insulating and conductive states can nucleate in close vicinity of a liner defect of a Dirac material. Conducting states have linear dispersion relation with a peculiar slowdown of the propagation velocity (Fig. \ref{fig:fig1} (d)) controlled by the matching properties induced by the defect. Renormalized propagation velocity follows the approximate relation $v_{eff} \approx v(1-\Lambda^2/2)$, for interfaces with $\Lambda \ll 1$. Protection from backscattering events due to the valley-momentum locking mechanism can possibly occur. These findings suggest that renormalization of the group velocity, which is usually attributed to the presence of many-body interactions, can emerge also when interaction effects in the bulk are negligible compared to the kinetic energy of quasiparticles. The slowdown of the group velocity, when controlled by an external parameter (such as the twist angle in bilayer graphene or the misorientation angle in graphene grain boundaries), allows the modulation of the Hamiltonian kinetic term which can eventually become comparable with the particle-particle interaction term. Under this condition, correlated states emerge. Interestingly, when a correlated state is established, a further renormalization of the group velocity can take place.\\

Discussions with R. De Luca and M. Salerno are gratefully acknowledged.


\begin{thebibliography}{99}
\bibitem{feynman82} R. P. Feynman, Int. J. Theor. Phys. \textbf{21}, 467 (1982).
\bibitem{fpu55} E. Fermi, J. Pasta, S. Ulam, \textit{Studies of non linear problems}, Los-Alamos internal report, document LA-1940 (1955).

\bibitem{quantumcomp96} A. Ekert and R. Jozsa, Rev. Mod. Phys. \textbf{68} (3), 733 (1996).
\bibitem{dwavepc16} V. S. Denchev, S. Boixo, S. V. Isakov, N. Ding, R. Babbush, V. Smelyanskiy, J. Martinis, and H. Neven, Phys. Rev. X \textbf{6}, 031015 (2016).

\bibitem{dassarma2011} S. Das Sarma, S. Adam, E. H. Hwang, and E. Rossi, Rev. Mod. Phys. \textbf{83}, 407 (2011).
\bibitem{polini13} M. Polini, F. Guinea, M. Lewenstein, H. C. Manoharan, and V. Pellegrini, Nature Nanotechnology \textbf{8}, 625 (2013).
\bibitem{kleintun} P. E. Allain and J. N. Fuchs, Eur. Phys. J. B \textbf{83}, 301 (2011).
\bibitem{rusin08}	T. M. Rusin and W. Zawadzki, Phys. Rev. B \textbf{78}, (2008).
\bibitem{tikhonenko09} F. V. Tikhonenko, A. A. Kozikov, A. K. Savchenko, and R. V. Gorbachev, Phys. Rev. Lett. \textbf{103}, (2009).
\bibitem{ostrovsky08}	P. M. Ostrovsky, I. V. Gornyi, and A. D. Mirlin, Phys. Rev. B \textbf{77}, (2008).
\bibitem{cheianov07}	V. V. Cheianov, V. Fal'ko, and B. L. Altshuler, Science \textbf{315}, 1252 (2007).
\bibitem{cao18a}	Y. Cao, V. Fatemi, A. Demir, S. Fang, S. L. Tomarken, J. Y. Luo, J. D. Sanchez-Yamagishi, K. Watanabe, T. Taniguchi, E. Kaxiras, R. C. Ashoori, and P. Jarillo-Herrero, Nature \textbf{556}, 80 (2018).
\bibitem{cao18b} Y. Cao, V. Fatemi, S. Fang, K. Watanabe, T. Taniguchi, E. Kaxiras, and P. Jarillo-Herrero, Nature \textbf{556}, 43 (2018).
\bibitem{chen19} G. Chen, A. L. Sharpe, P. Gallagher, I. T. Rosen, E. J. Fox, L. Jiang, B. Lyu, H. Li, K. Watanabe, T. Taniguchi, J. Jung, Z. Shi, D. Goldhaber-Gordon, Y. Zhang, and F. Wang, Nature (2019); https://doi.org/10.1038/s41586-019-1393-y.
\bibitem{magicangles19} G. Tarnopolsky, A. J. Kruchkov, and A. Vishwanath, Phys. Rev. Lett. \textbf{122}, 106405 (2019).

\bibitem{gbstates16} A. Luican-Mayer, J. E. Barrios-Vargas, J. T. Falkenberg, G. Autès, A. W. Cummings, D. Soriano, G. Li, M. Brandbyge, O. V. Yazyev, S. Roche, and E. Y. Andrei, 2D Materials \textbf{3}, 031005 (2016).

\bibitem{giamarchibook} T. Giamarchi, \textit{Quantum Physics in One Dimension} (Clarendon; Oxford University Press, Oxford: New York, 2004).
\bibitem{note} Throughout this work symbols $\mathbb{I}$ and $\sigma_{x,y,z}$ are used to indicate the $2 \times 2$ identity matrix and the Pauli matrices, respectively. The symbol $\otimes$ is used to indicate the Kronecker product of matrices. The notation $p_{x}=-i\hbar \partial_{x}$ and $p_{y}=-i\hbar \partial_{y}$ is introduced for quantum operators associated with the linear momentum components. Complex conjugation operator is denoted by $\mathcal{K}$, while $\textbf{A}^t$ denotes the transpose of $\textbf{A}$.

\bibitem{trgraphene} H. Suzuura and T. Ando, Phys. Rev. Lett. \textbf{89}, 266603 (2002).
\bibitem{matchingm} F. Romeo and A. Di Bartolomeo, Materials \textbf{11}, 1660 (2018).
\bibitem{beenakker06} C. W. J. Beenakker, Phys. Rev. Lett. \textbf{97}, 067007 (2006).
\bibitem{photonicgraphene} T. Ochiai and M. Onoda, Phys. Rev. B \textbf{80}, 155103 (2009).

\end{thebibliography}
\end{document}